\documentclass[11pt]{article}

\usepackage{amsmath}  \usepackage{amsfonts} \usepackage{psfrag}
\usepackage{graphicx}

\newcommand{\nn}{\nonumber} \newcommand{\ra}{\rightarrow}

\newcommand{\ddpsi}{\ddot{\psi}}
\newcommand{\dpsi}{\dot{\psi}}
\newcommand{\be}{\begin{equation}}
\newcommand{\ee}{\end{equation}}
\newcommand{\ba}{\begin{eqnarray}}
\newcommand{\ea}{\end{eqnarray}}

\newcommand{\va}{\varepsilon}
\newcommand{\s}{\sigma}
\newcommand{\Tr}{{\rm Tr}}
\newcommand{\dH}{\dot{H}}

\newcommand{\tF}{\tilde{F}} \newcommand{\tR}{\tilde{R}}
\newcommand{\cE}{\mathcal{E}}\newcommand{\cH}{\mathcal{H}}
\newcommand{\vf}{\varphi}

\newcommand{\cL}{\mathcal{L}}
\newcommand{\cD}{\mathcal{D}}
\newcommand{\vinf}{V_{\mathrm{infl}}}
\newcommand{\Ne}{N_{e-\mathrm{folds}}}\newcommand{\Mpl}{M_{\mathrm{pl}}}
\newcommand{\mg}{\mathrm{g}}

\begin{document}
\title{\bf{HYM-flation: Yang-Mills cosmology with Horndeski coupling}}

\author{E. Davydov\thanks{davydov@theor.jinr.ru}~~and  D. Gal'tsov\thanks{galtsov@phys.msu.ru}\\
\small{\it{ $^*$ Bogoliubov Laboratory of Theoretical Physics,
JINR, 141980, Dubna, Moscow region, Russia, }}\\
\small{\it{$^\dag$  Faculty of Physics of the Moscow State
University, 119899, Moscow, Russia }}}\maketitle

\begin{abstract}
We propose new mechanism for inflation using classical $SU(2)$
Yang-Mills (YM) homogeneous and isotropic  field non-minimally
coupled to gravity via Horndeski prescription. This is the unique
generally and gauge covariant ghost-free YM theory  with the
curvature-dependent action leading to second-order gravity and
Yang-Mills field equations. We show that its solution space contains
de Sitter boundary to which the trajectories are attracted for some
finite time, ensuring the robust inflation with a graceful exit. The
theory can be generalized to include the Higgs field leading  to
two-steps inflationary scenario, in which  the Planck-scale
YM-generated inflation naturally prepares the desired initial
conditions for the GUT-scale Higgs inflation.
\end{abstract}

\section{Introduction}
The 2015 Planck's collaboration release \cite{Planck15}  confirmed
that inflation had occurred at relatively low energy scale,
$10^{15}\div 10^{16}$~GeV, with almost absent non-Gaussianity, and
small tensor fluctuations being.  A variety of phenomenological
models with a scalar inflaton  slowly rolling down in a flat
potential describes well the  data. These include Starobinsky $R^2$
model~\cite{Starobinsky_R2}, Higgs inflation~\cite{Higgs_Inflation}
and some other traditional models demanding tuning of parameters at
the classical level. The tuning, however, is not protected from
large quantum corrections, so various attempts were undertaken to
find symmetries underlying the desired flatness of the potential
making inflation ``natural''. One possibility, based on scalar
fields only, invokes  models with hidden conformal or shift
symmetry~\cite{Conformal_attractors,Coupled_attractors,cosm_attractors,Shift_symmetry}.
The simplest such model~\cite{Conformal_attractors} contains in the
Jordan frame the following combination of the Einstein term and the
conformally coupled scalar field term
\begin{equation}\label{00}
L_{\rm conf}\sim\left(1-\frac{\phi^2}{6}\right)\frac{R}{2},
\end{equation}
vanishing at the boundary $\phi^2=6$. In the Einstein frame this
leads to flattening of the scalar potential $V(\phi)\ra
V(\sqrt{6}\tanh(\vf/\sqrt{6})$, which results in  de Sitter solution
as $\vf\ra\infty$. Higgs inflation is particularly attractive since
it identifies the inflaton with some known field. Note that this
mechanism invokes the non-minimal coupling of Higgs to gravity via
the curvature scalar.

The  GUT-scale conformal inflation, however, raises the problem of
initial conditions, discussed already in the early days of inflation
scenario~\cite{New_inflation}, and recently reconsidered again in
the modern setting~\cite{Initial_Conditions1,Initial_Conditions2}.
One of its solutions  is a stage of preliminary inflation starting
at Planck scale and driving the inflaton to the plateau of the
GUT-scale observed inflation. The second inflaton is  usually taken
as another scalar field. Here we suggest to use for this purpose the
vector YM field which is generically present in gauge/supergravity
theories, so one does not need to introduce the second scalar by
hand. To realize such a scenario one has to assume, likewise in the
Higgs case, the non-minimal coupling of YM to gravity using the
Horndeski prescription~\cite{Horndeski_vector}.

Recall that the SU(2) Yang-Mills field has an isotropic and
homogeneous  mode and satisfies (in the case of the standard YM
lagrangian) the conformal equation of state $p=\epsilon/3$, thus
mimicking the hot Universe~\cite{Galtsov:1991un}. It was studied
in 90-ies both in classical and quantum minisuperspace settings
\cite{Moni} and Euclidean quantum gravity \cite{Donets:1992ck}.
Recently this idea was revived in anticipation of the future
precise measurements of the primordial gravitational waves imprint
on  the cosmic microwave background
(CMB)~\cite{Bielefeld:2014nza}. It was found that
the perturbations of YM and tensor gravitational perturbations mix
together, leading to difference in the
evolution of right- and left-polarized gravitational waves (parity
violation) which can become testable soon.

During past decade various modifications of the standard YM action
breaking the conformal symmetry  in a way consistent with the
Standard Model and its extensions were introduced. Replacing the
YM lagrangian by the Born-Infeld string motivated lagrangian, e.
g., leads to an equation of state interpolating between that of
the string gas and the photon gas $-\epsilon/3<p<\epsilon/3$, but
this is insufficient for inflation \cite{BI}. Phenomenologically,
it was noticed that the lagrangian generically depending on two
invariants $\cL(f,g),\; f=F^{a}_{\mu\nu}F^{a\mu\nu},\;g={\tilde
F}^{a}_{\mu\nu} F^{a\mu\nu},$ leads to desired inflationary
equation of state if the dependence  on   $g$ is non-linear, i.e.
$\partial^2 \cL/\partial g^2\neq 0$~\cite{Gauge_inflation2}. A
particular model of this type with the quadratic $g^2-$term in the
lagrangian was called ``gauge-flation'' and received a lot of
attention~\cite{Gauge_inflation}. But physical origin of such a
term, which, moreover, must enter with a  large coefficient, is
somewhat obscure.

Conformal symmetry is also broken once interaction of the YM field
with the scalar fields is introduced. This is what happens in the
full gauge theories. It was shown that the YM-Higgs model with the
complex doublet Higgs leads to a kind of hybrid inflation
scenario,~\cite{Gal'tsov:2010dd} in which dynamics of the Higgs
field is modulated by the YM component (this model was recently
revived in \cite{Rinaldi:2015iza}). Other typical interactions of
YM fields include dilaton~\cite{Fuzfa:2005qn} and axion. The
latter option attracted much attention as implementing the idea of
``naturalness''~\cite{Natural} and became known as
``chromo-natural'' inflation~\cite{Adshead:2012kp}.

The next class of models, closer to the present one, consists in
exploiting the non-minimal coupling of the inflaton to gravity,
which was applied to Higgs field under the name of Higgs inflation
\cite{Higgs_Inflation,Higgs_criticality}. The idea was also
applied to the YM field possibly together with modifying gravity
lagrangian
\cite{Balakin,Bamba:2008xa,Banijamali:2011ep,YM_nonminimal2}. The
potential danger of curvature-modified gravity is the emergence of
higher derivative field equations plagued with the Ostrogradski
ghosts. In attempts to avoid ghosts, new ideas associated with
massive gravity and/or galileons were invoked \cite{Galileons,
Galileons_Lovelock}. General classes of couplings of  vector
fields to gravity involving curvature tensor couplings whose
equations of motion does not contain ghosts were found by
Horndeski
 \cite{Horndeski_vector}. Initially the non-minimal vector coupling
to gravity  was introduced as the  extension of the Maxwell theory
in curved space which preserves the second order equations of
motion, admits the energy-momentum and charge conservation laws, and
reduces to Maxwell theory in the flat space  limit. Later it was
revealed that Lovelock gravity~\cite{Lovelock},  galileon
models~\cite{Galileons} and Horndeski theory~\cite{Horndeski_scalar,
Horndeski_vector} are strongly
 interrelated~\cite{Galileons_Lovelock}. Using the Abelian
 vector fields in cosmology
\cite{Horndeski_Abelian1,Horndeski_Abelian2} leads  either to
isotropy or gauge invariance problems. So here we consider the
unique case free from these complications which was not discussed
before: the lowest order Hordeski coupling of the SU(2) YM field
to the dual Riemann tensor. This model contains only one extra
parameter of the dimension of mass which turns out to be the
Hubble constant of the de Sitter stage in this model. We
demonstrate that de Sitter solution is the boundary of the
solution space which attracts a large set of trajectories, keeps
them for some finite time and then relaxes to the hot universe
state.

\section{Non-minimal coupling of vector field to gravity}
General gauge-invariant curvature-dependent action quadratic in the
vector field strength $F_{\mu\nu}$  and linear in the curvature can
be written in the form
\begin{equation}\label{01}
S_{\rm RF}=\int
\mathcal{R}^{\alpha\beta\mu\nu}F_{\alpha\beta}F_{\mu\nu}\sqrt{-g}
d^4x\,,
\end{equation}
where the {\it susceptibility} tensor
$\mathcal{R}^{\alpha\beta\mu\nu}$ has the same index permutation
symmetries as the Riemann tensor. It can be presented as the linear
combination
\begin{equation}\label{0a}
    \mathcal{R}^{\alpha\beta\mu\nu}=  4q_2R^{[\alpha[\mu}g^{\nu]\beta]}+
    q_1 R g^{\alpha[\mu}g^{\nu]\beta} -q_3R^{\alpha\beta\mu\nu}
    \,,
\end{equation}
where $R^{\alpha\beta\mu\nu}$ is the Riemann tensor, $R^{\alpha\mu}$
is the Ricci tensor,   $R$ is the scalar curvature, and the brackets
$[\;]$ mean an alternation over indices with the factor $1/2$. Such
a structure is typical for the one-loop corrections to the Maxwell
action in  curved  space QED~\cite{Nonminimal_scattering}, where the
coefficients $q_1,\; q_2,\;q_3$ have certain particular values. Here
we consider this action as phenomenological, but subject to some
theoretical restrictions. The field $F_{\mu\nu}$ in (\ref{01}) can
be either Abelian, or non-Abelian, in which case we will  use the
matrix notation,
 \be
A_\mu=A^a_\mu T_a\,,\quad F_{\mu\nu}=F_{\mu\nu}^a T_a=2\nabla_{[\mu}
A_{\nu]}+\left[A_\mu,\,A_\nu\right]\,,\ee assuming the SU(2) gauge
group  \be \left[T_a,T_b\right]=\va_{ab}^{\;\;\;c} T_c\,,\quad
\Tr(T_a T_b)=\frac12\delta_{ab}\,,
  \ee
and adding  the trace operator $\tR$ before the lagrangian.

For generic coefficients $q_1,\, q_2\,,q_3$ the resulting theory
contains higher derivatives generating extra degrees of freedom
which are plagued with Ostrogradski ghosts. The unique
curvature-dependent coupling leading to the ghost-free theory was
found by Horndeski~\cite{Horndeski_vector}. It corresponds to
$q_1=q_2=q_3$, in which case the susceptibility tensor reduces to
the double-dual Riemann tensor:
\begin{equation}\label{0}
 \tR^{\alpha\beta\gamma\delta}=\frac14
\epsilon^{\alpha\beta\mu\nu}R_{\mu\nu\rho\sigma}\,\epsilon^{\rho\sigma\gamma\delta}\,,
\end{equation}
where the Levi-Civita tensors contain suitable $\sqrt{-g}$ factors.
This tensor satisfies the Bianchi identity
 \be
\nabla_\alpha \tR^{\alpha\beta\mu\nu}=0\,,
  \ee
which is crucial for making the theory ghost-free.

  Note that the Horndeski action can be  written
in two equivalent forms: \be \label{Ho} S_{\rm H}= \Tr\int
{\tR}^{\alpha\beta\mu\nu} F_{\alpha\beta} F_{\mu\nu} \sqrt{-g}\,
d^4x=\Tr\int
{R}^{\alpha\beta\mu\nu}\tF_{\alpha\beta}\tF_{\mu\nu}\sqrt{-g}\, d^4x
\,,
 \ee using the dual field tensors
$\tF^{\alpha\beta}\equiv\frac12\epsilon^{\alpha\beta\mu\nu}F_{\mu\nu}$.
This structure is reminiscent of the Gauss-Bonnet lagrangian
\begin{equation}\label{0b}
    L_{\rm GB}=-\tR^{\alpha\beta\mu\nu}R_{\alpha\beta\mu\nu}=R^2-
    4R^{\alpha\beta}R_{\alpha\beta}+
    R^{\alpha\beta\mu\nu}R_{\alpha\beta\mu\nu}\,,
\end{equation}
from which it can be obtained replacing the Riemann tensor by the
product of two field tensors. This is not accidental: the Horndeski
action can be  derived from the higher-dimensional Gauss-Bonnet
theory by dimensional reduction~\cite{YM_from_GB}. It is worth
noting that the vector Horndeski lagrangian  in four dimensions is
much simpler than the scalar Horndeski one~\cite{Horndeski_scalar}
which was widely used recently in attempts to improve the simplest
non-minimal Higgs inflation model~\cite{Horndeski_scalar_infl}.

Using the variation of the Riemann tensor
 \be \delta
R_{\alpha\beta\mu\nu}=R^{\rho}_{[\beta\mu\nu}\delta
g_{\rho\alpha]}+
    2  \nabla_{[\mu}\nabla_{[\beta}\delta
    g_{\alpha]\nu]}
 \ee
and the Bianchi identity for the YM field
  \be D_\mu
\tF^{\mu\nu}=0\,,
 \ee
where the gauge covariant derivative is introduced,
 \be D_\mu
F_{\alpha\beta}\equiv \nabla_\mu F_{\alpha\beta}+[A_\mu,
F_{\alpha\beta}]\,,
 \ee
one can write the variation of the action (\ref{Ho}) over the metric
in the form
 \be
 \begin{split}
\frac{\delta S_{\rm H}}{\sqrt{-g}\;\delta g_{\rho\s}}= -2\Tr\left(
 -\frac14 g^{\rho\sigma}\tR^{\alpha\beta\mu\nu} F_{ \alpha\beta}
 F_{\mu\nu}\right.
&+F^{(\rho}_{\;\; \beta}\tR^{\sigma)\beta\mu\nu}F_{ \mu\nu}
+R_{\alpha\beta}\tF^{ \alpha\rho}\tF^{ \beta\sigma}\\
&\left.+D_\beta\tF^{\alpha\rho}D_\alpha\tF^{\beta\sigma}
+F_{\alpha\beta}\!\left[\tF^{\alpha\rho},\tF^{ \beta\sigma}\right]
\right)\,,
\end{split}
  \ee
in which the absence of higher derivatives is manifest. On the
contrary, one can notice the presence of the cubic term in $F$.

We conclude this section with brief review of the earlier proposals
to use non-minimally coupled vector fields in cosmology. The
slow-roll inflation model with (generic) non-minimally coupled
Maxwell field was suggested in~\cite{Maxwell_nonminimal}, yet
suffering the issue of anisotropy. The latter has  been evaded in
the non-Abelian case~\cite{YM_nonminimal1, YM_nonminimal2}, however
these earlier models either were loosing gauge invariance or
contained ghosts. The Abelian vector model with Horndeski coupling
both gauge invariant and without ghosts was investigated
in~\cite{Horndeski_Abelian1}, but this model was unable to provide
de Sitter solutions unless the cosmological constant was added by
hand~\cite{Horndeski_Abelian2}. The Yang-Mills-Higgs cosmology with
general non-minimal coupling~(\ref{0a}) was examined
in~\cite{Balakin}, where de Sitter solutions were found, but no
detailed investigation  of inflation was undertaken.

In what follows we will study  inflationary solutions in the
ghost-free SU(2) Horndeski Yang-Mills  model   showing that robust
inflation emerges because of general property of the Horndseki
coupling, which closely resembles the coupling of the scalar field
used in the models of conformal attractors. It turns out that the YM
non-linearity is crucial for possibility of this scenario: it is
impossible in the Maxwell case.

\section{HYM cosmology}
It is convenient to rescale coordinates and the YM potential  as
follows: $x^\mu\ra x^\mu/(\mg\Mpl)$, $A_\mu\ra \Mpl A_\mu$, where
$\Mpl=1/\sqrt{8\pi G}=2.435\times 10^{18}$~GeV is the Planck mass,
and $\mg$ is a gauge coupling constant. Then we choose the units
$\mg\Mpl=1$. Actually, in most gauge theories the coupling constant
is of the order of unity, so in what follows we assume
$\mg\Mpl\sim\Mpl$.

Adding the Einstein term and the standard YM term to the Horndeski
action we obtain the total action
 \be\label{2}
S_{\rm HYM}=\frac{1}{2\mg^2}\int\left(R-\Tr\left(
F_{\mu\nu}\Phi^{\mu\nu}
    \right)\right) \sqrt{-g}\,d^4 x\,,
 \ee
where the ``induction'' tensor is introduced with the coupling
$\mu^{-2}$ (dimensionless in the rescaled quantities):
 \be
 \Phi^{\mu\nu}=F^{\mu\nu}+\frac1{2\mu^2}\tR^{\mu\nu\lambda\tau}F_{\lambda\tau}\,.
 \ee

Even before passing to the Friedmann metrics, one can notice the
following fundamental property of the HYM action: its matter part
vanishes in de Sitter space with some curvature radius. Indeed,
adjusting the Hubble parameter of de Sitter to be \be \label{Hmu}
H^2=\mu^2\,,\ee one finds
\begin{equation}\label{2a}
    \tR^{\alpha\beta\mu\nu}=-\mu^2\left(g^{\alpha\mu}g^{\beta\nu}-
    g^{\alpha\nu}g^{\beta\mu}\right)\,,
\end{equation}
in which case $\Phi^{\mu\nu}=0$. Obviously this has to be the
boundary of the physical domain, beyond which we would get the
phantom YM field (such an option is not considered here). We will
see that the inflationary solutions are attracted to the phantom
boundary, but do not cross it.

It is worth noting that the existence of the de Sitter boundary  is
encountered in the more general non-minimal theory~(\ref{01}) too
(see~\cite{Balakin}). In this case we get
\begin{equation}\label{2a2}
    \mathcal{R}^{\alpha\beta\mu\nu}=-H^2[6(q_1-q_2)+q_3]
    \left(g^{\alpha\mu}g^{\beta\nu}-g^{\alpha\nu}g^{\beta\mu}\right)\,,
\end{equation}
so in de Sitter space with the Hubble parameter, satisfying
$(6(q_1-q_2)+q_3)H^2=\mu^2$  instead of (\ref{Hmu}), the induction
tensor also vanishes.

The Einstein equations of the HYM theory can be written in the usual
form
   \be
R_{\mu\nu}-\frac12 g_{\mu\nu}R=T_{\mu\nu}\,,
   \ee
where the effective energy-momentum tensor reads
 \be
 \begin{split}
T_{\mu\nu}=&2\Tr\left(F_{(\mu\beta}\Phi_{\nu)}^{\;\;\beta}
-\frac14 g_{\mu\nu} F_{\alpha\beta}\Phi^{\alpha\beta}\right)\\
&+\frac{1}{\mu^2}\Tr\left(R_{\alpha\beta}\tF^{\alpha}_{\;\mu}\tF^{\beta}_{\;\nu}
+D_{\beta}\tF^{\alpha}_{\;\mu}D_{\alpha}\tF^{\beta}_{\;\nu}+
F_{\alpha\beta}\!\left[\tF^{\alpha}_{\;\mu},\tF^{\beta}_{\;\nu}\right]\right)\,.
\end{split}
 \ee
The equations of motion for YM field are simply
\begin{equation}\label{3b}
    D_\nu \Phi^{\mu\nu}  =0\,.
\end{equation}

Now we pass to the homogeneous and isotropic cosmology, restricting
for simplicity by the spatially flat metric:
\begin{equation}\label{4}
     ds^2=-N^2 dt^2+a^2\left[dr^2+r^2
 (d\theta^2+\sin^2\theta d\varphi^2)\right]\,.
\end{equation}
The YM matrix-valued one-form can be written in certain gauge in
terms of a single function $\psi(t)$ (for more general gauges and
any spatial curvature see \cite{YM_cosm,BI}):
\begin{equation}\label{4a}
    A=a\psi [T_r dr+r (T_\theta d\theta+ T_\varphi\sin\theta
    d\varphi)]\,,
\end{equation}
showing that the direction of ${\bf A}$ in the color space coincides
with the space direction. Choosing the proper time gauge $N=1$ and
introducing the `electric' and `magnetic' effective fields:
$\cE\equiv\dpsi+H\psi$, $\cH\equiv\psi^2$ we can present the
standard YM Lagrangian  in the Maxwell form:
\begin{equation}\label{4d}
-\frac14 F_{\mu\nu}^a
F^{a\mu\nu}=\frac32\left(\cE^2-\cH^2\right)\,.
\end{equation}
The effective one-dimensional Lagrangian of the $SU(2)$ HYM
model~(\ref{2}) then reads:
\begin{equation}\label{5x}
L=6H^2+3\dot{H}\left(1+\frac{\cH^2}{2\mu^2}\right)+\frac32
\left(1-\frac{H^2}{\mu^2}\right)\left(\cE^2-\cH^2\right)\,,
\end{equation}
where one can notice the  factor $1-H^2/\mu^2$ in the gauge sector,
indicating on the de Sitter boundary described above.

It is easy to check that the stress-energy tensor is diagonal and
isotropic, with  the energy density and pressure
\begin{eqnarray}
  &\rho_g=&\frac32\left(\dpsi^2+2H\psi\dpsi+H^2\psi^2+\psi^4\right)-\nn\\
 & &\quad\frac{3}{2\mu^2}\left[H^2(3\dpsi^2+3H^2\psi^2+2\psi^4)+
 2H\psi\dpsi(3H^2+2\psi^2)\right]\,,\label{7b}\\
  &p_g=&\frac12\left(\dpsi^2+2H\psi\dpsi+H^2\psi^2+\psi^4\right)+
  \frac{1}{2\mu^2}\left[3\dpsi^2(3H^2+4\psi^2)+2H\psi\dpsi(7H^2+8\psi^2)\right.\nn\\
  &&\quad\left.+H^2\psi^2(5H^2+2\psi^2)+4\ddpsi(H\dpsi+H^2\psi+\psi^3)+
  2\dH(\dpsi^2+4H\psi\dpsi+3H^2\psi^2)\right].\label{7d}
\end{eqnarray}
In the limit of vanishing coupling, $\mu\ra\infty$, the system
represents radiation with the equation of state $p_g=\rho_g/3$.

The dynamics of the system is governed by Friedmann equations,
\begin{equation}\label{6}
  H^2=\frac{\rho_g}{3}\,,\quad
  \dH+H^2=-\frac16(\rho_g+3p_g)\,,
\end{equation}
and the gauge field equation,
\begin{equation}\label{7}
   \left[1-\frac{H^2}{\mu^2}\right]\left(\dpsi+H\psi\right)\dot{\mathstrut}+
   2\left[1-\frac{\dot{H}+ H^2}{\mu^2}\right]\left(H\dpsi+H^2\psi+\psi^3\right)=0\,.
\end{equation}
An important characteristic of the system~(\ref{6}--\ref{7}) is the
determinant of the matrix of coefficients before the  derivatives
$\dH,\,\ddpsi$\,:
\begin{equation}\label{6c2}
\cD\equiv
\left[1-\frac{H^2}{\mu^2}\right]\left(1+\frac{1}{2\mu^2}\left[(\dpsi+H\psi)^2-
2\psi^4\right]\right)+\frac{2}{\mu^4}\left(H\dpsi+H^2\psi+\psi^3\right)^2\,.
\end{equation}
When this quantity vanishes, the solution meets the singularity. One
can show that the boundary $H^2=\mu^2$ separates the domain of
non-singular solutions from that of singular ones. Consider a
solution crossing the boundary at the moment $t=t_1$, so that
$H^2(t_1)=\mu^2,\, \dot{H}(t_1)\neq 0$. Then the gauge field
equation~(\ref{7}) implies that the expression in the round brackets
in the second term vanishes: $H\dpsi+H^2\psi+\psi^3=0$\,. This
implies vanishing of  the determinant $\cD$, indicating the
singularity. Thus the non-singular trajectories should not cross the
boundary $H^2=\mu^2\,$. The physical trajectory must be non-singular
and reach the flat space asymptotic: $H=\psi=\dpsi=0$, which implies
$\cD\ra 1$. Therefore, physical initial states should reside in the
following domain of the phase space:
\begin{equation}\label{63}
    \mathbb{D}_{\mathrm{phys}}=\{H^2<\mu^2\}\cap \{\cD>0\}\,,
\end{equation}
to which the flat space asymptotic belongs. Note that with $H<\mu$
the sign of the kinetic term of the Lagrangian~(\ref{5x}) remains
positive. Thus the boundary $H^2=\mu^2$ also preserves the system
from falling into the phantom state.

\section{HYM-flation temporary attractor}
Contrary to earlier negative verdict concerning the Abelian
Horndeski-Maxwell cosmology ~\cite{Horndeski_Abelian1,Vector_cosm}
we would like to show that non-linearity of the YM theory makes the
proposed HYM model much more promising. Our claim is that  \emph{de
Sitter solution $H^2=\mu^2$ is an inflationary attractor in
non-Abelian Horndeski model; robust inflation emerges without
fine-tuning of parameters or initial conditions.}

 To show this we first observe that the YM Eq. (\ref{7}) is
satisfied if $H^2=\mu^2,\; \dH=0$.  Then one can solve the first
Friedmann equation in~(\ref{6}) as a quadratic equation in $\dpsi$:
\begin{equation}\label{71a}
    \dpsi_{\pm}=-\frac{1}{\mu}\left(\psi^3+\mu^2\psi\pm\sqrt{\psi^6+(3/2)\psi^4
    \mu^2-\mu^4}\right)\,.
\end{equation}
Since only two of the three equations~(\ref{6},~\ref{7}) are
independent, the remaining second Friedmann equation will also be
satisfied with $H=\mu$ and $\dpsi$ given by~(\ref{71a}).
Obviously, the square root in~(\ref{71a}) is real only if the YM
function is above the critical value $\psi>\psi_{\rm cr}$,
satisfying $\psi_{\rm cr}^6+(3/2)\psi_{\rm cr}^4 \mu^2-\mu^4=0$.
This solution is possible due to the YM non-linearity, which
manifests itself in presence of the terms $\psi^6,\,\psi^4$ under
the square root. In the Abelian case the first Friedmann equation
would imply $\mu^2=-\cE^2$ for the ansatz $H^2=\mu^2,\; \dH=0$.

The critical value, $\psi_{cr}$, is proportional to $\mu^{2/3}$ for
$\mu\ll 1$, and to $\mu^{1/2}$ for $\mu\gg 1$. The large field
limit, $\psi\gg \max(\mu,\mu^{2/3})$, corresponds to the dominance
of $\psi^6$ under the square root in~(\ref{71a}) and always
satisfies $\psi\gg\psi_{cr}$, as well. Then the two branches of the
solution~(\ref{71a}) simplify and can be easily integrated:
\begin{eqnarray}\label{72}
    &&\dpsi_{+}\simeq -\frac{2\psi^3}{\mu}\quad \Rightarrow \quad \psi_{+}\simeq
    \sqrt{\frac{\mu}{4(t-t_0)}}\,,\\
 &&\dpsi_{-}\simeq -\frac{\mu\psi}{4} \quad \Rightarrow \quad \psi_{-}\simeq \psi_0
    e^{-\mu t/4}\,.\label{73}
\end{eqnarray}
The first branch corresponds to dominance of the kinetic term,
$\cE\gg\cH$, while in second case the YM potential prevails,
$\cH\gg\cE$.

Consider now small deviations $(\delta
H,\,\delta\psi,\,\delta\dpsi)$ from these solutions and compute the
eigenvalues of the corresponding linearized systems. These values
can be viewed as  local Lyapunov exponents~\cite{Lyapunov_Exponents}
which describe the growth rate of the deviations   in a given mode
(note that the solutions (\ref{72}) are not the stationary point of
the system). The result reads:
\begin{eqnarray}
    &\psi_{+}:&\frac{12\psi^2}{\mu},\,\frac{2\sqrt{15}\psi^2}{\mu},\,
    -\frac{2\sqrt{15}\psi^2}{\mu}\,,\nn\\
 &\psi_{-}:&-2\mu,\,-\frac{\mu}{4},\,-\frac{5\mu}{4}\,.\label{74}
\end{eqnarray}
From this we deduce that $\psi_{+}$ is an unstable singular
solution, while $\psi_{-}$ describes the mode which is stable for
some period of time. For this reasons we call the solution
$\psi_{-}$ an `inflationary attractor',  though strictly speaking it
is not an attractor in the sense of the theory of dynamical system.
The universe filled with the supercritical YM condensate at high
density, $\psi\gg\max(\mu,\mu^{2/3})$, experiences inflation with
$H=\mu$ and $\psi\propto e^{-\mu t/4}$. As expansion is going on,
the gauge condensate monotonously decays. Eventually, when $\psi$
drops below $\psi_{cr}$, the de Sitter stage ends and transition
takes place to the universe filled with radiation.

\subsection{Constant-roll regime}
One can use numerics to explore behavior of solutions with different
initial data demonstrating explicitly that their choice within the
substantial region of the physical domain
$\mathbb{D}_{\mathrm{phys}}$ (\ref{63}) ensures qualitatively
similar behavior. Namely, after some time the solution gets
attracted to the de Sitter stage of finite duration. To show
robustness of this process we choose zero initial value for the YM
field $\psi_i=0$. Then one can easily see that two parameters
defining $\mathbb{D}_{\mathrm{phys}}$,
\begin{equation}\label{75}
    H^2_i=\mu^2\frac{ \dpsi^2_i}{3\dpsi^2_i+2\mu^2}\,,\quad
    \mathcal{D}_i=\frac{3\dpsi^4_i+3\mu^2\dpsi^2_i+2\mu^4}{\mu^2(3\dpsi_i^2+2\mu^2)}\,,
\end{equation}
satisfy the desired conditions $0\leq H^2_i<\mu^2/3$, $\cD_i\geq 1$
for any $\dpsi_i$.

\begin{figure}[tb]
\begin{center}
\begin{minipage}[t]{0.48\linewidth}
\hbox to\linewidth{\hss%
\psfrag{1}{{$\mu t$}} \psfrag{2}{\Large{$\frac{\psi}{\mu}$}}
\psfrag{3}{{$\psi\propto e^{-\mu t/4}$}}
  \includegraphics[width=0.95\linewidth,height=0.7\linewidth]{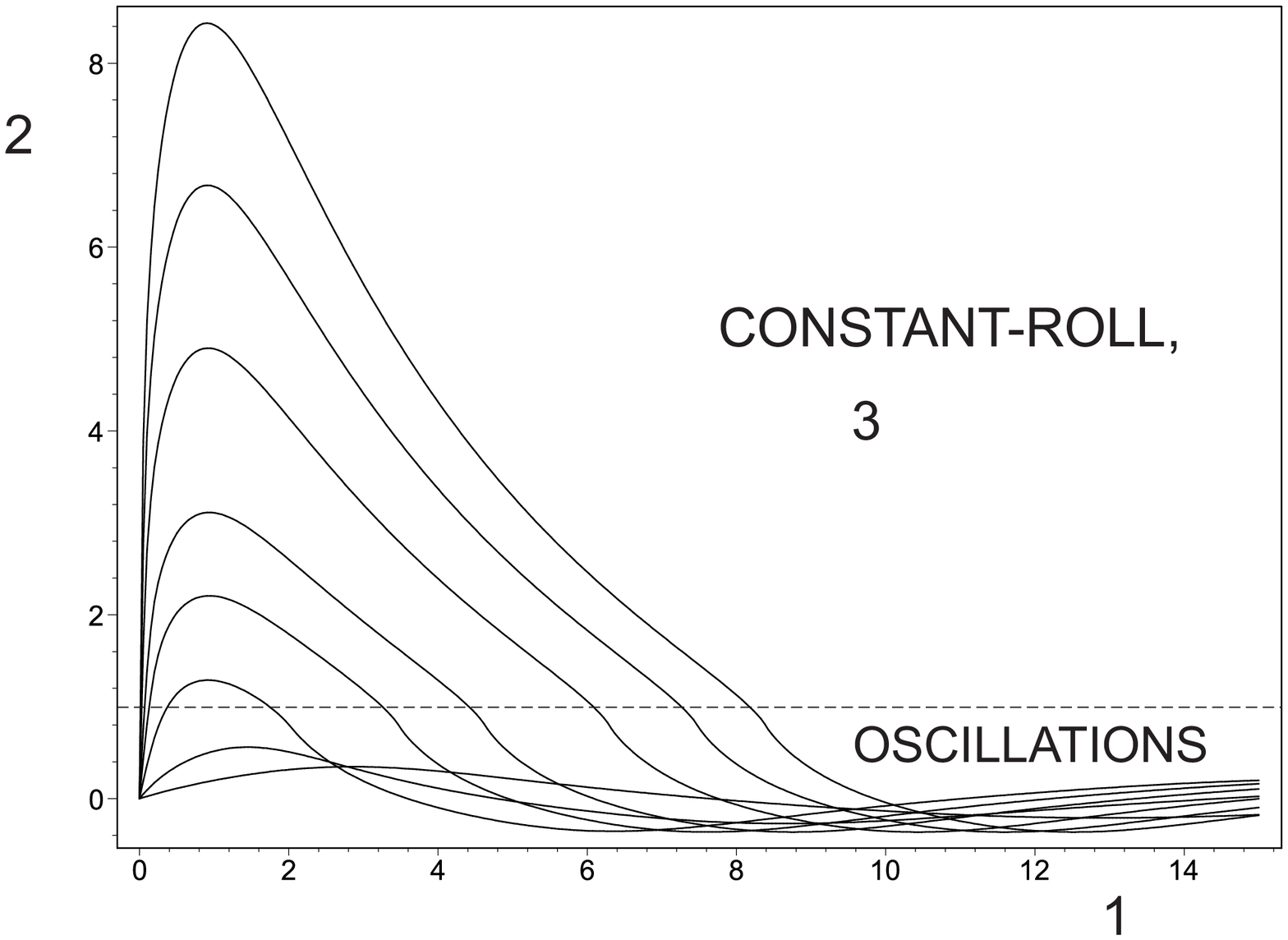}
\hss} \caption{\small   The solutions for gauge field, $\psi$. The
coupling cut-off scale, $\mu$, separates exponentially decaying
and oscillating modes.} \label{F1}
\end{minipage}
\hfill
\begin{minipage}[t]{0.48\linewidth}
\hbox to\linewidth{\hss%
\psfrag{1}{{$\mu t$}} \psfrag{2}{\Large{$\frac{H}{\mu}$}}
  \includegraphics[width=0.95\linewidth,height=0.7\linewidth]{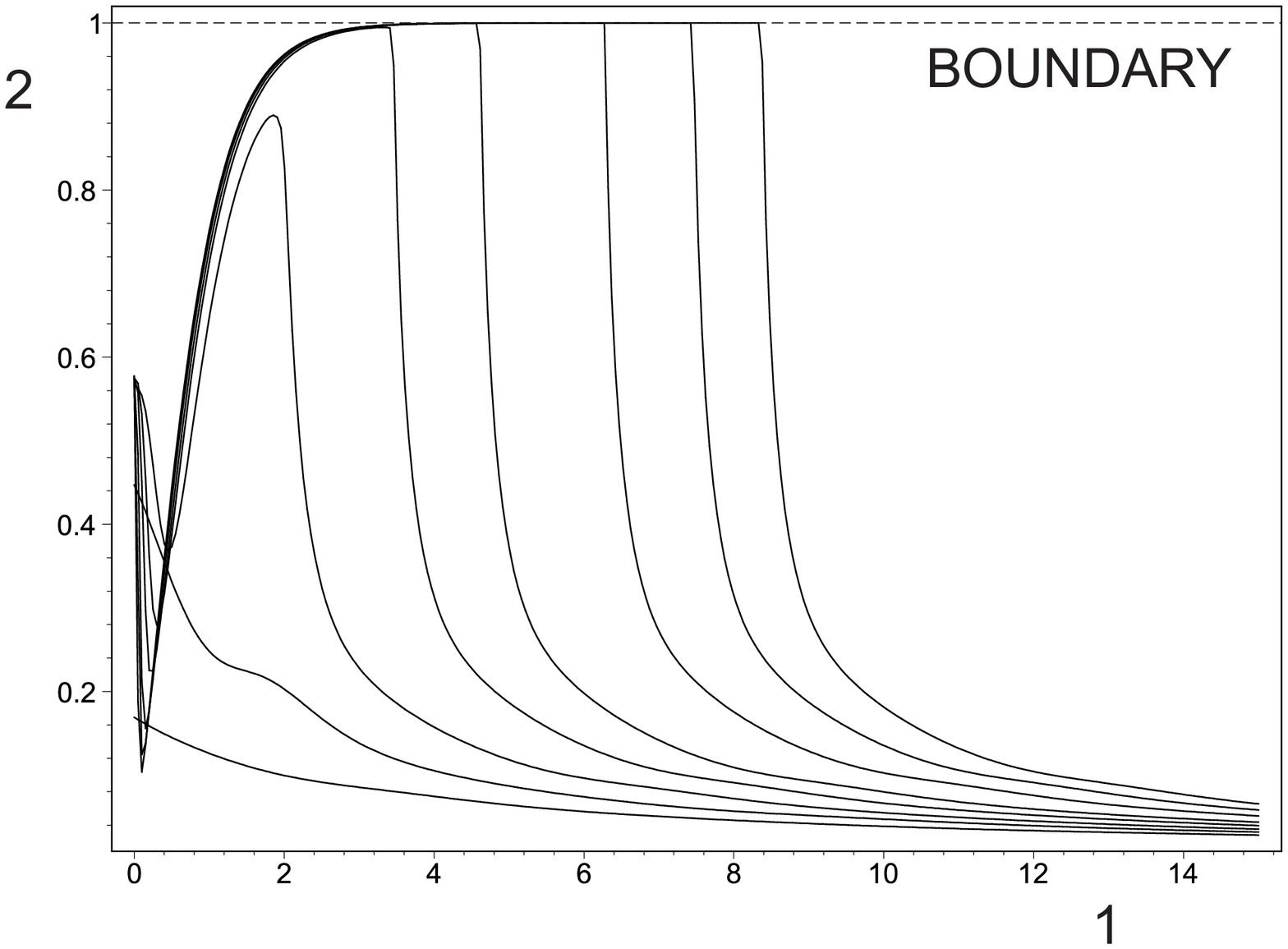}
\hss} \caption{\small   The solutions for Hubble parameter, $H$.
The trajectories are temporarily attracted to the boundary
$H^2=\mu^2$. Then transition to radiation dominated universe,
$H=1/2t$, occurs.} \label{F2}
\end{minipage}
\end{center}
\end{figure}

On the Fig.~(\ref{F1},\,\ref{F2}) we present evolution of the YM
function and the Hubble parameter  starting from the initial states
$\psi_i=0,\,\dpsi_i/\mu^2=0..100$.   Such initial states correspond
to domination of the kinetic energy, so the Hubble parameter rapidly
decreases. Therefore, initially the system resides in the phase
space region distant from the inflationary attractor and even moves
further away from it. Nonetheless, soon after, the trajectories
starting with $\dpsi_i\gg\mu^2$  become attracted to the  boundary
$H=\mu$, signalling the onset of inflation. The finite duration of
this stage is ensured by a subsequent exponential  decay of the YM
function along with the exponential growth of the scale factor. Once
the field value eventually falls below the critical value $\psi_{\rm
cr}$, the effect of the non-minimal coupling becomes negligible.
Then the system undergoes  transition to the radiation dominated
stage $H\ra 1/(2t)$, corresponding to an oscillating  YM function.

Despite the fact that dynamical system~(\ref{6},\,\ref{7}) looks
rather complicated, the physics  behind it resembles much the very
first model of chaotic inflation~\cite{Chaotic_inflation} with
$\lambda\vf^4$ potential, which is just an oscillator with the
Hubble friction. In our case the self-interaction of non-Abelian
gauge fields plays crucial role, generating the $\psi^4$ potential.
The cosmology in the non-Abelian Horndeski theory can be viewed as
an oscillator \emph{with the Hubble friction amplified by
non-minimal coupling to gravity}, provided the initial state belongs
to physical domain, $\mathbb{D}_{\mathrm{phys}}$. The precise
initial conditions are totally irrelevant, only the energy of
oscillations matters. If this energy  allows the field to climb high
enough above the non-minimal coupling scale, $\psi \gg\mu $, the
system at maximal deviation will be inevitably attracted by the
solution $H=\mu$, and the amplified Hubble friction will impose the
subsequent steady downhill motion. Therefore the domain of
inflationary initial states can be presented as
\begin{equation}\label{76}
    \mathbb{D}_{\mathrm{infl}}=\mathbb{D}_{\mathrm{phys}}
    \cap\{\max{(\psi^4,\,\dpsi^2)}\gg\mu^4\}\,.
\end{equation}

Comparing our model with scalar inflation  with a monomial potential
$V(\vf)\propto \vf^n$ one observes the following important
difference. In the scalar slow-roll case the Hubble parameter
rapidly grows with  increasing field value, $H^2\sim V(\vf)$. In the
HYM model, the $H^2$ dependence is flattened near the boundary of
the phase space:  qualitatively  $H^2\sim f(\tanh\psi/\mu)$. This
resembles the {\em conformal attractors}, in which case the
effective potential is flattened near the phase space boundary, so
that $H^2\sim V(\tanh\vf/\sqrt{6})$. In the present model the
potential $\psi^4$ is quite steep, while the Hubble friction is
constant during the inflationary stage. This is why one observes the
{\em constant-roll} motion~\cite{Constant_roll}, when
$\ddot{\psi}/H\dpsi=n=\mathrm{const}$. In our case $n=-1/4$, while
the slow-roll motion corresponds to $n\simeq 0$ (and $n=-3$ is
`ultra slow-roll').

\begin{figure}[tb]
\hbox to\linewidth{\hss%
\psfrag{0}{\Huge{$0$}} \psfrag{1}{\Huge{$\psi$}}
\psfrag{2}{\Huge{$\mu$}} \psfrag{3}{\huge{$\lambda^{-\frac14}$}}
\psfrag{4}{\Huge{$1$}} \psfrag{5}{\huge{$\lambda^{-\frac16}$}}
\psfrag{6}{\Huge{$\frac{\lambda}{4}\vf^4$}}
\psfrag{7}{\Huge{$\frac12\psi^4$}}
\psfrag{8}{\Huge{$H\sim\sqrt{\lambda}\vf^2$}}\psfrag{9}
{\Huge{$H=\mu=\mathrm{const}$}}
\psfrag{z}{\Huge{$\vf$}}
  \resizebox{10cm}{7cm}{\includegraphics{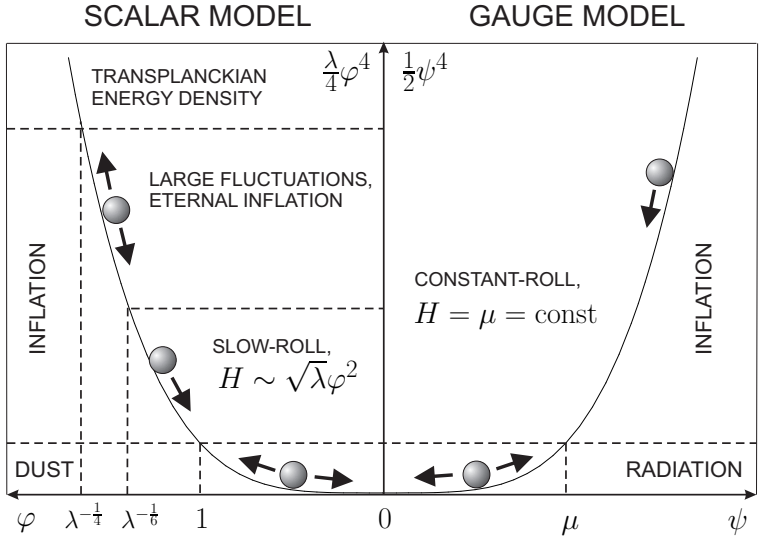}}
\hss} \caption{\small  The schematic description of the standard
slow-roll inflation with the scalar inflaton (left panel) and the
constant-roll HYM-flation with the non-minimally coupled SU(2)
vector field (right panel). The HYM-flation occurs near the phase
space boundary resulting in the constant Hubble friction and
consequently a substantially different behavior. } \label{F3}
\end{figure}

The peculiar feature of the constant-roll inflation is the absence
of eternal inflation. Indeed, quantum fluctuations should be of the
order $\delta\psi\approx H/2\pi\approx\mu/6$. But according to
classical motion, for the time interval $t=H^{-1}=\mu^{-1}$, the YM
field decreases by the value $\Delta\psi\approx\psi
\,(1-e^{-1/4})\approx\psi/5$. During most of the inflation stage,
one has $\psi\gg\mu$, therefore quantum fluctuations provide just a
minor correction to the classical motion: $\Delta\psi\gg
\delta\psi$. Contrary to the standard chaotic inflation
scenario~\cite{Eternal}, here the Hubble parameter (and therefore
the amplitude of quantum fluctuations) remains constant, while the
potential $\psi^4$ becomes steeper as the field value grows. Hence
the field can not climb up due to quantum fluctuations. Only at the
exit from the inflation stage one has $\psi\sim \mu$, and the
quantum fluctuations may become comparable with the classical
values. This can probably generate the large scale structure of the
universe, but it can not lead to an eternal self-reproduction. It is
a matter of opinion whether the eternal inflation is
objectionable~\cite{Selfreproduction}, or not. But anyway, our
inflationary scenario seems to be free of it.

To summarize the results, on the Fig.~(\ref{F3}) we schematically
compare the chaotic HYM-flation emerging in the Horndeski
vector-tensor theory with the $\psi^4$ self-coupling potential of
the YM field, and the chaotic inflation in the minimally coupled
scalar field theory with the potential $\lambda\vf^4$.

\section{HYM-flation as pre-inflation}

During the constant-roll motion $\psi$ exponentially decreases with
the rate $-\mu/4$, while the scale factor grows with the rate $\mu$.
So the number of $e$-folds gained by the scale factor at the moment
$t_e$ is just the number of $e$-folds lost by  $\psi$ with a factor
of four:
\begin{equation}\label{7h}
    \Ne=\mu\,(t_e-t_i)\simeq 4\ln{\psi_i/\psi_e}\,.
\end{equation}
Interpreting HYM-flation as the observed inflation, we will be faced
with the problem of perturbations, however. If one considers $\mu$
as the scale parameter of the observed inflation, so that $\mu\sim
10^{-6} \Mpl$, then $\psi_e$, being slightly larger than $\mu$ for
the inflationary solution, must be of the order $10^{-6}\div 10^{-5}
\Mpl$. Naturally choosing the initial conditions at the Planckian
scale, $\psi_i\sim \Mpl$, one has $\Ne\approx 46\div 55\,$. Though
technically this can be regarded satisfactory as the minimally
required number of $e$-folds, one can barely obtain the observed
perturbation spectrum in such a model. An accurate analysis of
perturbation spectra in YM cosmology is rather involved (see
calculations in the context of the
``gauge-flation''~\cite{Gauge_inflation}), but qualitative estimates
do not seem to be in favor of the HYM-flation as the model of the
observed inflation. Indeed, the amplitude of fluctuations is
inversely proportional to $\dpsi$, which exponentially decreases
with time. Hence, the power spectrum should be rather blue-tilted,
what contradicts to the Planck's data.

 Adopting a view  that the observed
inflation should occur at the GUT scale and be described by the
scalar field slow-rolling in the plateau potential, we can suggest
the Planck scale HYM accelerated expansion  as the {\em
pre-inflation}. It could help to solve the issue of initial
conditions for the observed inflation. Recall that this problem was
raised long ago in the context of the so-called new
inflation~\cite{New_inflation} scenario. That time it was alleviated
with invention of the chaotic inflation~\cite{Chaotic_inflation} in
which the de Sitter expansion starts immediately after Planck era.
Nowadays, observational data give preference to the low scale
inflation with the plateau potential, during which the energy
density is nearly $\rho\simeq\vinf\sim 10^{-10}$. Thus a
pre-inflationary evolution of the universe from some initial state
with Planck energy $\rho\sim 1$ is claimed again.

Note that the problem of an initial excess of kinetic energy, $K\sim
1\gg \vinf\sim 10^{-10}$, is not a critical one. Kinetic energy
density drops as fast as $K\propto a^{-6}$, while the value of the
inflaton field changes not significantly during the pre-inflation
epoch~\cite{Initial_Conditions1}. So, after a period of the
post-planckian expansion, the inflaton can likely be found on a
plateau, where the potential energy dominates. Of course, the closed
universe with the Planck-size volume filled with non-inflating
matter will collapse long before the low scale inflation could
start. But this objection can be evaded arguing that universe was
born non-closed.

The problem of inhomogeneities is more
thorny~\cite{Initial_Conditions2}. If the universe in the
pre-inflationary epoch was dominated by the kinetic energy or
radiation,  the scale factor would grow  as $a\propto t^{1/3}$ or
$a\propto t^{1/2}$. The observed slow-roll inflation with the
plateau potential $\vinf\sim 10^{-10}$ have to start  when the
kinetic energy density decreases by nearly eleven or twelve orders
of magnitude, which requires the scale factor to gain roughly two
orders of magnitude. This would take  quite a long time,
$t_{\mathrm{infl}}\sim 10^4\div 10^5$. But in the decelerating
universe, $a\propto t^{n}$, $n<1$, the cosmological horizon grows
with time:
\begin{equation}\label{9}
    d_{\mathrm{hor}}(t_1,t_2)=a(t_1)\int_{t_1}^{t_2}\frac{dt}{a(t)}\simeq
    \frac{n}{H(t_1)(1-n)}\left(\frac{t_2}{t_1}\right)^{1-n}\,.
\end{equation}
If the universe was born at Planck time, $t_1=t_{Pl}=1$, with Planck
energy density, $H(t_1)=1/\sqrt{3}$, the distance at which the
initial inhomogeneities could spread during till the start of
inflation  $t_2=t_{\mathrm{infl}}$ then would be
\begin{equation}
    d_{\mathrm{hor}}(t_{Pl},t_{\mathrm{infl}})\sim 10^2\div 10^3.
\end{equation}
So, either the Universe was created homogeneous in a domain
containing millions or billions of the identical Planck volumes
(implausible), or some special topology has to be assumed
~\cite{Chaotic_mixing, Linde_after_Planck}, or  the preliminary
chaotic-type inflation~\cite{Open_universe, Fast_Roll} must be
introduced which started immediately after the Planck era and ended
before the stage of observed inflation.

HYM-flation as {\em pre-inflation} looks as the most economic way
to solve the problem of initial conditions for the GUT-scale
inflation. This does not require  the second scalar field, giving
job to gauge fields already present in the GUT or supergravity
models. Such a scenario thus may be viewed as an extension of the
Higgs inflation within the full gauge theory involving vector
fields. Let us take the natural value of coupling parameter
$\mu=1$ and assume that the universe was born  with the Planck
energy density, $\rho_i\simeq 1$. According to~(\ref{75}), for
vanishing initial field value, $\psi_i=0$, the energy density
$\rho_i=1$ requires $\dpsi_i^2\gg 1$. Conversely, with
$\dpsi_i^2\ll 1$, one has $\rho_i\ra 0$, which is unlikely for the
quantum creation process. Thus the universe with large probability
emerges in a state belonging to HYM-flation domain,
$\mathbb{D}_{\mathrm{infl}}$, for which $\dpsi_i^2\gg 1$. Of
course, this is a simplified picture which does not take into
account the non-vanishing initial gauge field value, the
contribution of other fields into the energy density and so on.
But, anyway, in the chaotic-like approach with randomly
distributed initial conditions there should be non-small
probability to find the universe (or some part of the universe) in
a state belonging to HYM-flation domain,
$\mathbb{D}_{\mathrm{infl}}$.

As was argued above, the issue of initial conditions for the low
scale inflation can be resolved if the scale factor grows by the
factor of $10^2\div 10^3$ during the pre-inflation. This provides a
large enough homogeneous patch, the core of which remains
homogeneous until the time when the low scale inflation can start.
In the non-Abelian Horndeski theory, according to Eq.~(\ref{7h}), it
is enough to have $\psi_i=4\div 6$ in order to get the robust
inflation with the desired gain of the scale factor.  Mention that
the initial domination of the potential term is not required: large
kinetic energy of oscillating motion in the quartic potential due to
the YM self-action will be transformed into the large potential
energy, and then inflation starts. This extended HYM-Higgs two-stage
inflationary model will be considered in more details in a separate
publication.

\section{Outlook}

Previous attempts  to construct models of inflation using YM fields
were based on some {\em ad hoc} assumptions about mechanism of the
conformal symmetry breaking~\cite{Gauge_inflation,
Gauge_inflation2}, and, as a consequence, an introduction into the
lagrangian of the new terms whose theoretical origin remained
obscure. Our present mechanism looks more natural, appealing to now
very popular non-minimal gravity couplings. Moreover, even within
the realm of the Horndeski-inspired models, the present one is
especially attractive by it simplicity and uniqueness. It has an
intrinsic de Sitter attractor which is manifest already at the level
of the lagrangian, the property which can hardly be underestimated.
Its second  crucial feature is the presence of the quartic
self-interaction term which was absent in the previous Abelian
non-minimal models, in which case stable inflation could not be
achieved~\cite{Balakin, Vector_cosm}. We have demonstrated that in
the HYM theory the corresponding homogeneous and isotropic cosmology
has a robust inflationary stage starting from a large variety of
initial data. During inflation, the YM field decays down to some
limiting value where the exponential expansion stops, ensuring a
natural graceful exit. This HYM-flationary  solution shares the
features of both the plateau inflation and the chaotic inflation
with quartic potential: the Hubble parameter is nearly constant (as
for the plateau potential), but the potential remains steep. This
results in a specific constant-roll motion, instead of the
slow-roll, and leads to absence of eternal inflation.

We propose this model on a role of the Planck-scale pre-inflation
preparing the initial conditions  for  the GUT-scale observed
inflation which can be  both incorporated into the
Yang-Mills-Higgs model with Horndeski coupling of the YM field to
gravity. Such model would combine advantages of the Higgs
conformal inflation and the preliminary chaotic-like inflation in
a very natural and economic way. Similar construction seems to be
possible in the context of the supergravity models where the
corresponding pair of the YM field and a scalar also can be found.

\textbf{Acknowledgments.} This work was supported by the Russian
Foundation for Fundamental Research under grant 14-02-01092 and by
the Joint Institute for Nuclear Research grant 15-302-02.

\end{document}